\documentclass[9pt,twocolumn,twoside]{optica}
\setboolean{shortarticle}{false}
\setboolean{minireview}{false}

\title{Efficient Third Harmonic Generation in Composite Aluminum Nitride / Silicon Nitride Microrings}

\author[1]{Joshua B. Surya}
\author[1]{Xiang Guo}
\author[1]{Chang-Ling Zou}
\author[1,*]{Hong X. Tang}

\affil[1]{Department of Electrical Engineering, Yale University, New Haven, Connecticut 06511, USA}

\affil[*]{Corresponding author: hong.tang@yale.edu}

\dates{Compiled \today}

\begin{abstract}
Aluminum nitride and silicon nitride have recently emerged as important nonlinear optical materials in integrated photonics respectively for their quadratic and cubic optical nonlinearity. A composite aluminum nitride and silicon nitride waveguide structure, if realized, will simultaneously allow highly efficient second and third harmonic generation on the same chip platform and therefore assists 2f-3f self-referenced frequency combs. On-chip third harmonic generation, being a higher order nonlinear optics effect, is more demanding than second harmonic generation due to the large frequency difference between the fundamental and third harmonic frequencies which implies large change of refractive indices and more strigent requirements on phase matching. In this work we demonstrate high-efficiency third harmonic generation in a high Q composite aluminum nitride / silicon nitide ring cavity. By carefully engineering the microring resonator geometry of the bilayer structure to optimize the quality factor, mode volume and modal overlap of the optical fields, we report a maximum conversion efficiency of $180\%\,\mathrm{W^{-2}}$. This composite photonic chip design provides a solution for efficient frequency conversion over a large wavelength span, broadband comb generation and self-referenced frequency combs. 
\end{abstract}

\begin{document}

\maketitle
\thispagestyle{fancy}
\ifthenelse{\boolean{shortarticle}}{\abscontent}{}

\section{Introduction}


\begin{figure*}[ht]
\centering
\includegraphics[width=\linewidth]{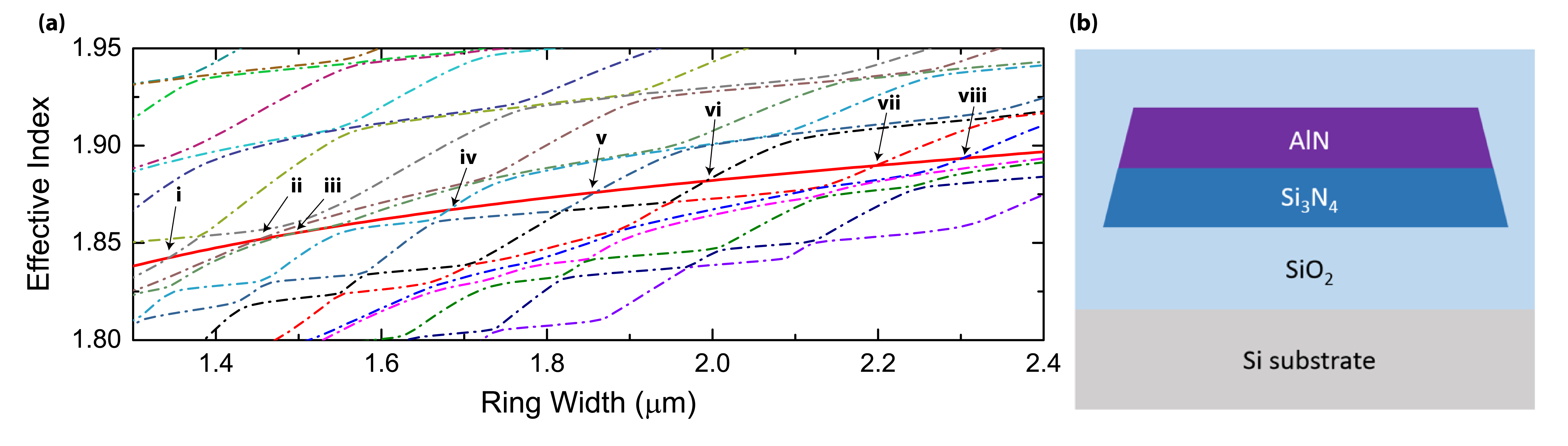}
\caption{\textbf{(a)} Effective index of the fundamental TE mode of the fundamental (solid red curve) and higher order modes of the third harmonic light (dotted curve) with varying ring widths. The intersecting points of the red curve with the other lines denote the ring widths that satisfy the phase matching. The phase-matched TE modes are marked with the corresponding labels from Table \ref{tab:modeprofile}. \textbf{(b)} Cross section of the composite waveguide. $\mathrm{330\thinspace nm}$ of AlN and $\mathrm{Si_{3}N_{4}}$ were grown on a $3.3\thinspace\mu \mathrm{m}$ thick $\mathrm{SiO_{2}}$ layer on top of an undoped $\mathrm{Si}$ substrate.}
\label{fig:effindex}
\end{figure*}
\begin{table*}[ht]
\centering
\caption{\bf Phase matched widths from simulations with the respective mode profiles and calculated overlap factor $\zeta$. }
\begin{tabular}{ccccccccc}
\hline
Width $\mu m$ & $1.34$ & $1.45$ & $1.49$ & $1.68$ & $1.86$ & $1.99$ & $2.20$ & $2.30$\\
\hline
\hfill & i. & ii. & iii. & iv. & v. & vi. & vii. & viii.\\
Simulated Mode Profile & \includegraphics[width=1.3cm]{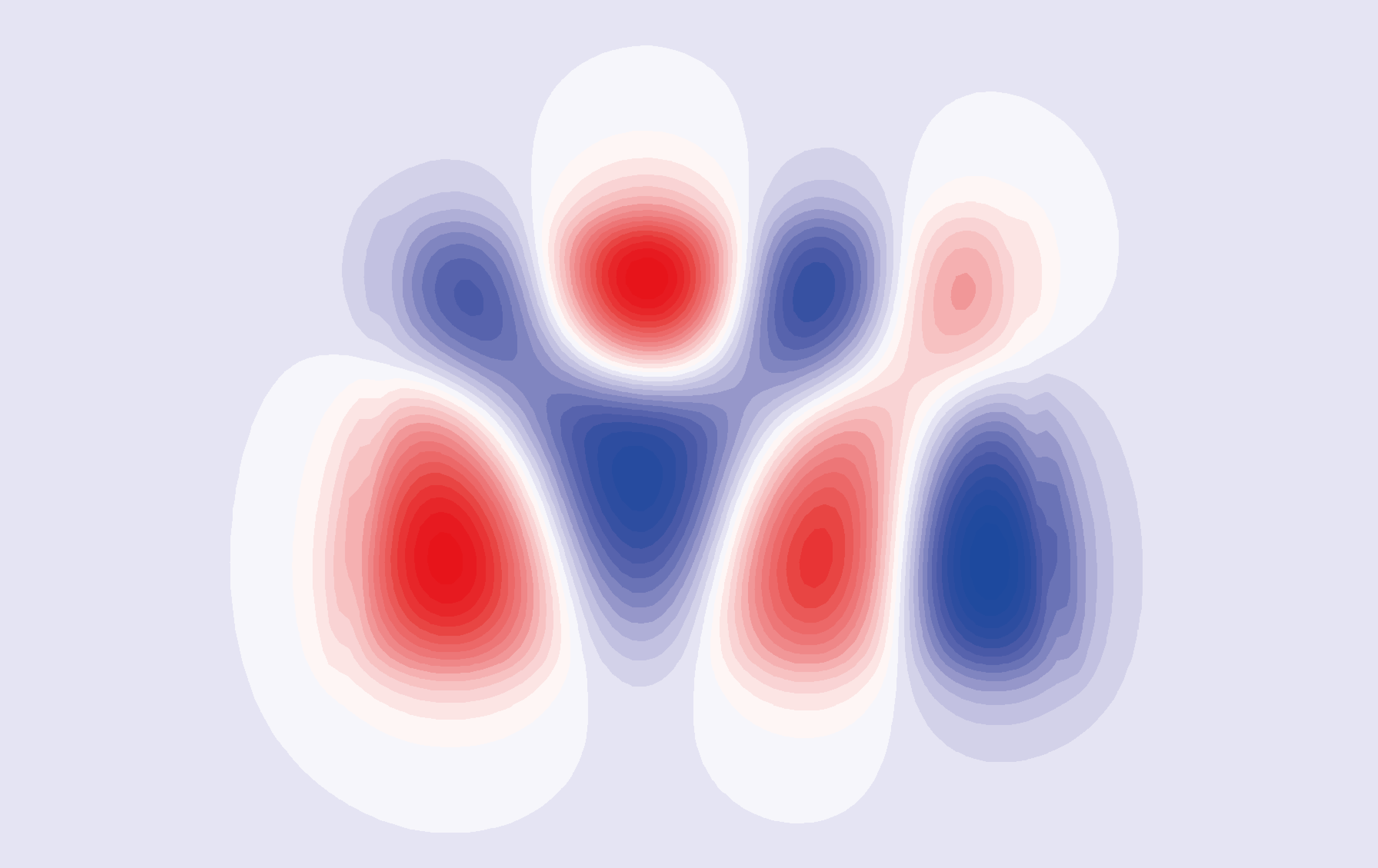} & \includegraphics[width=1.3cm]{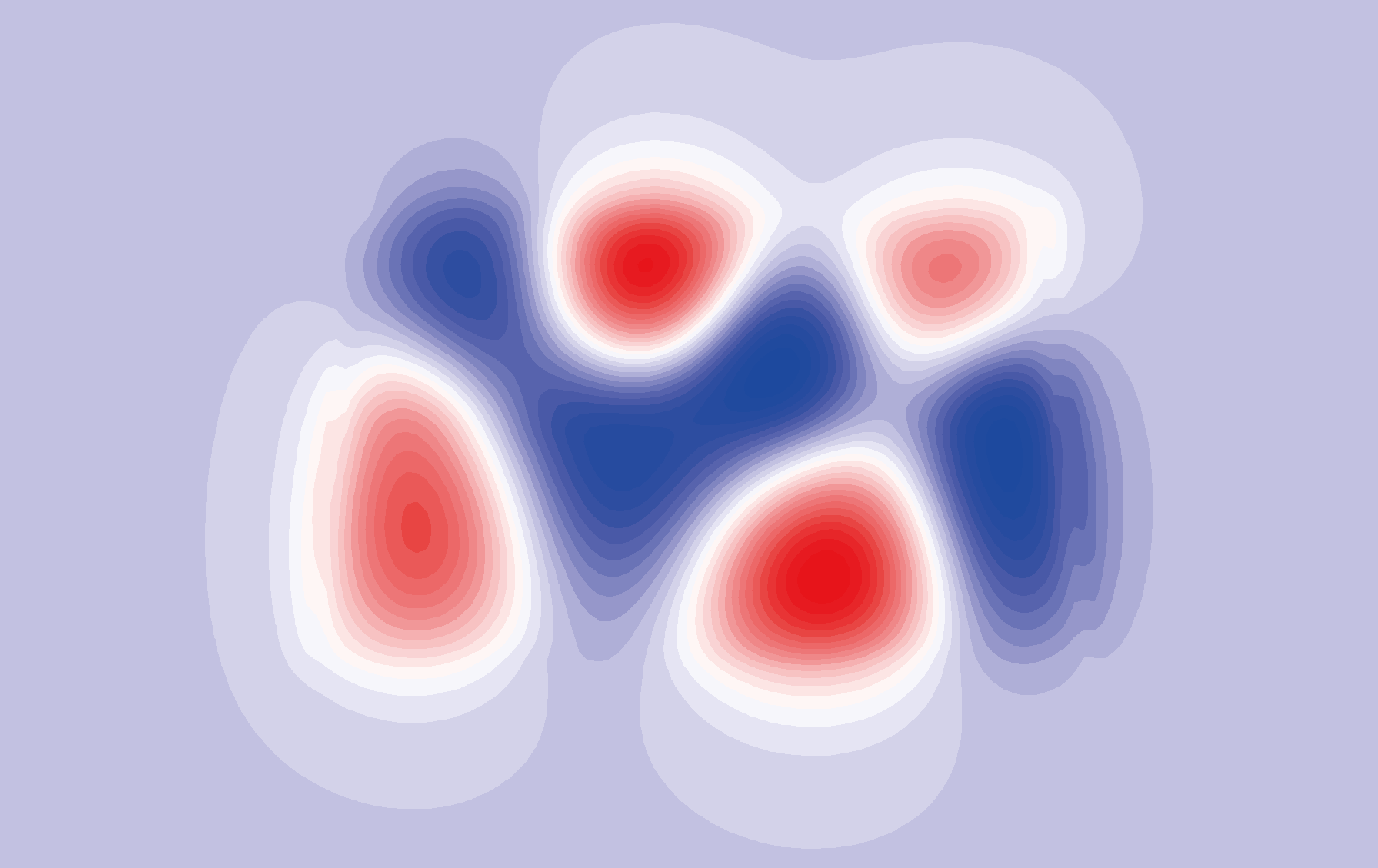} & \includegraphics[width=1.3cm]{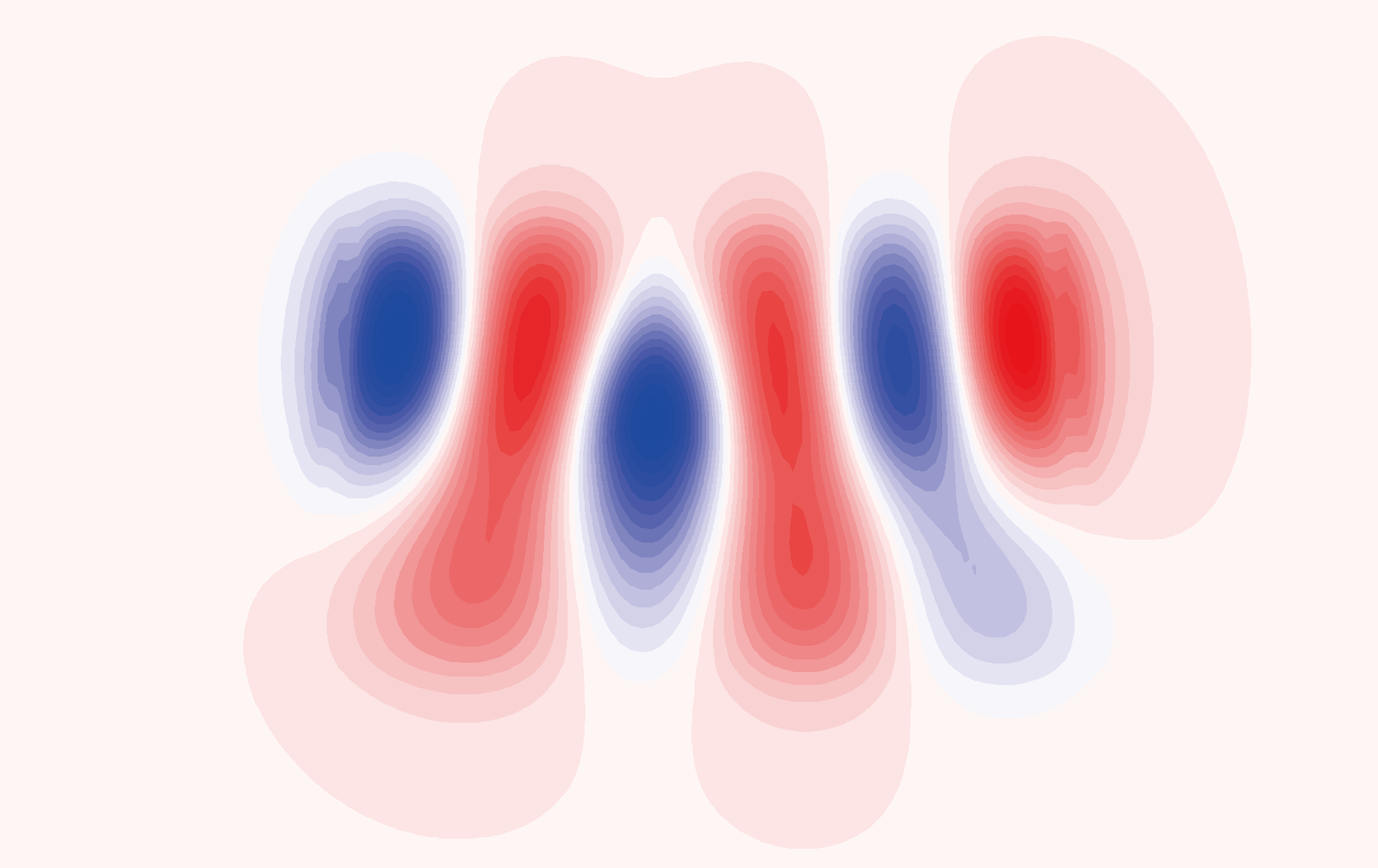} & \includegraphics[width=1.3cm]{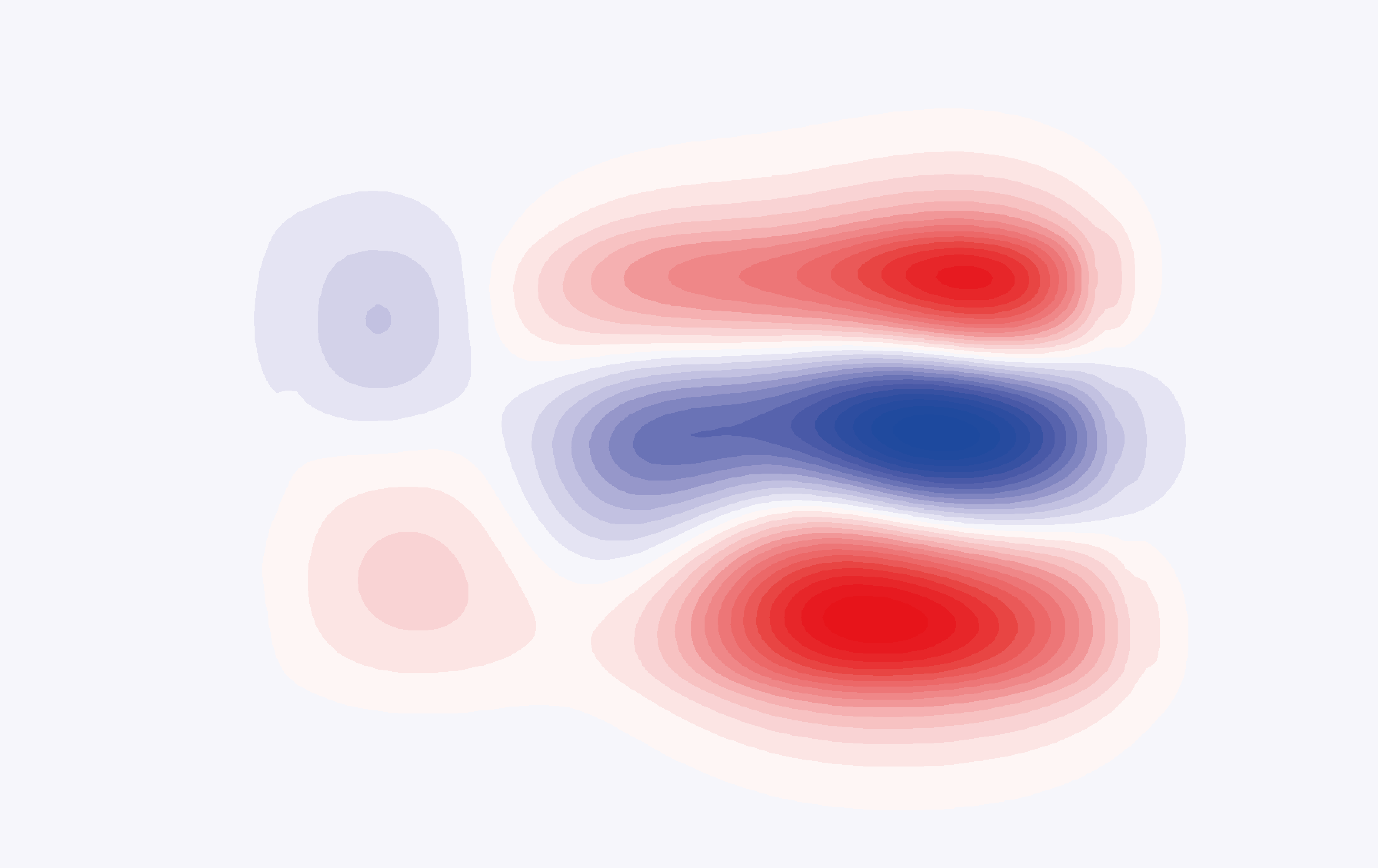} & \includegraphics[width=1.3cm]{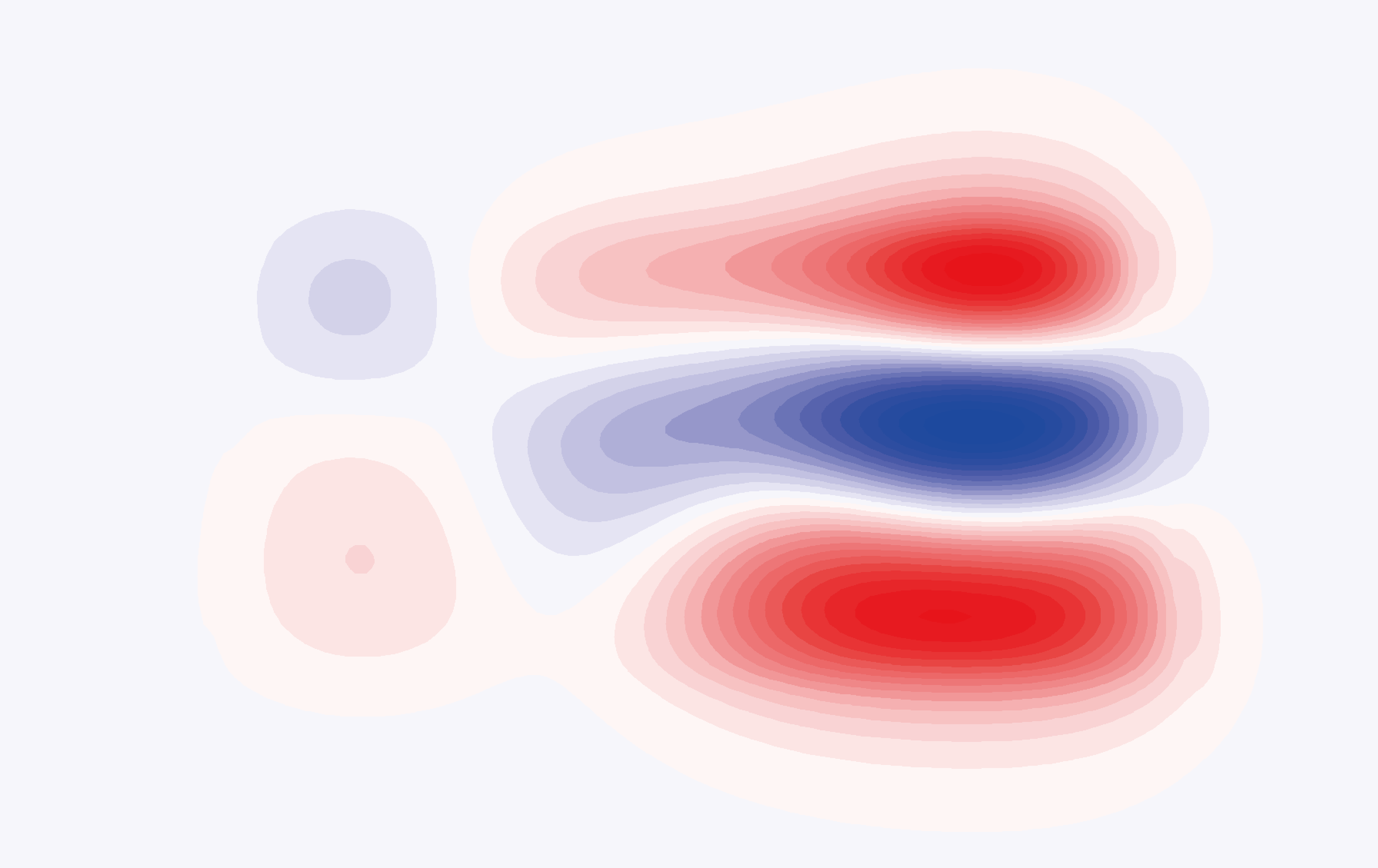} & \includegraphics[width=1.3cm]{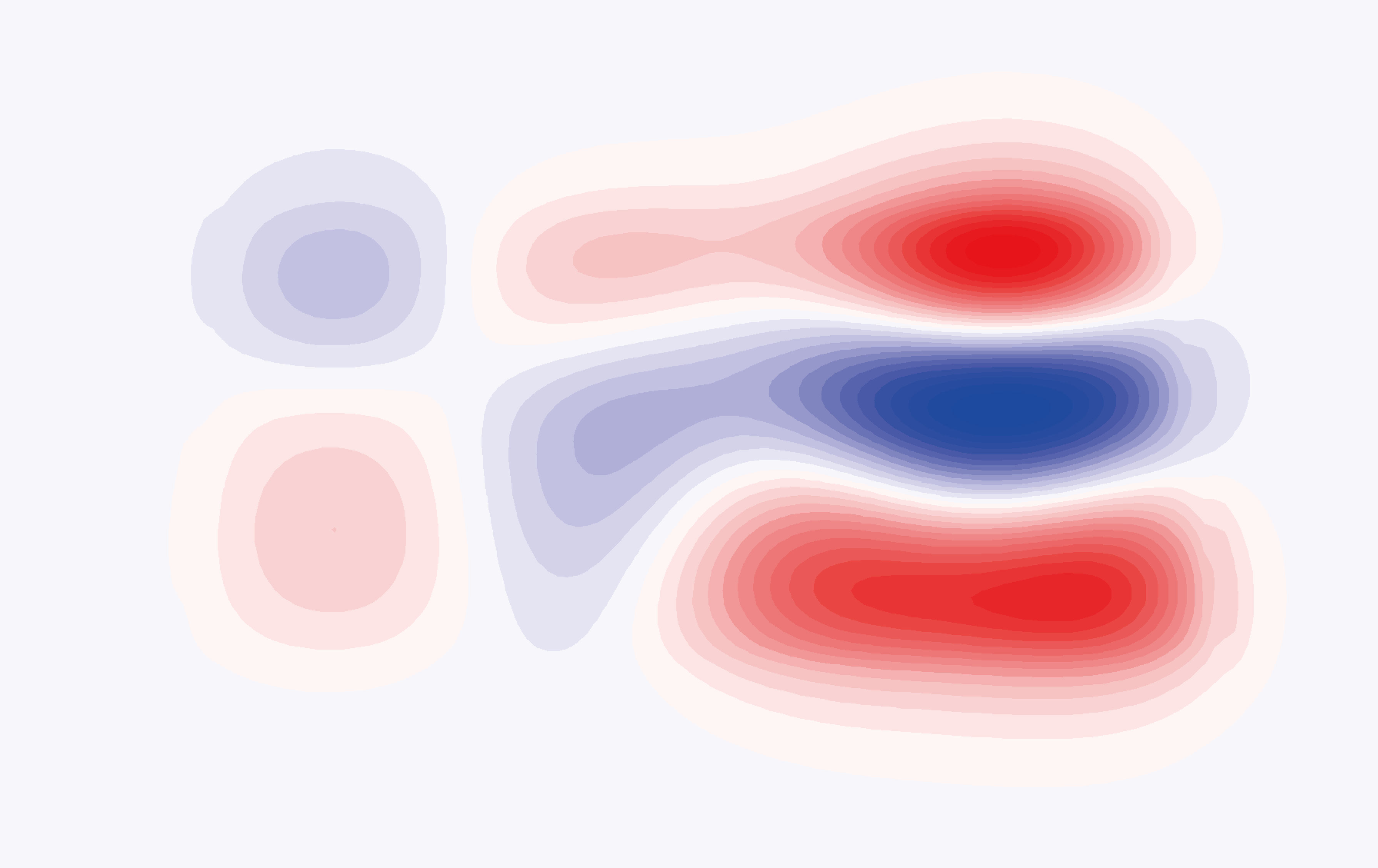} & \includegraphics[width=1.3cm]{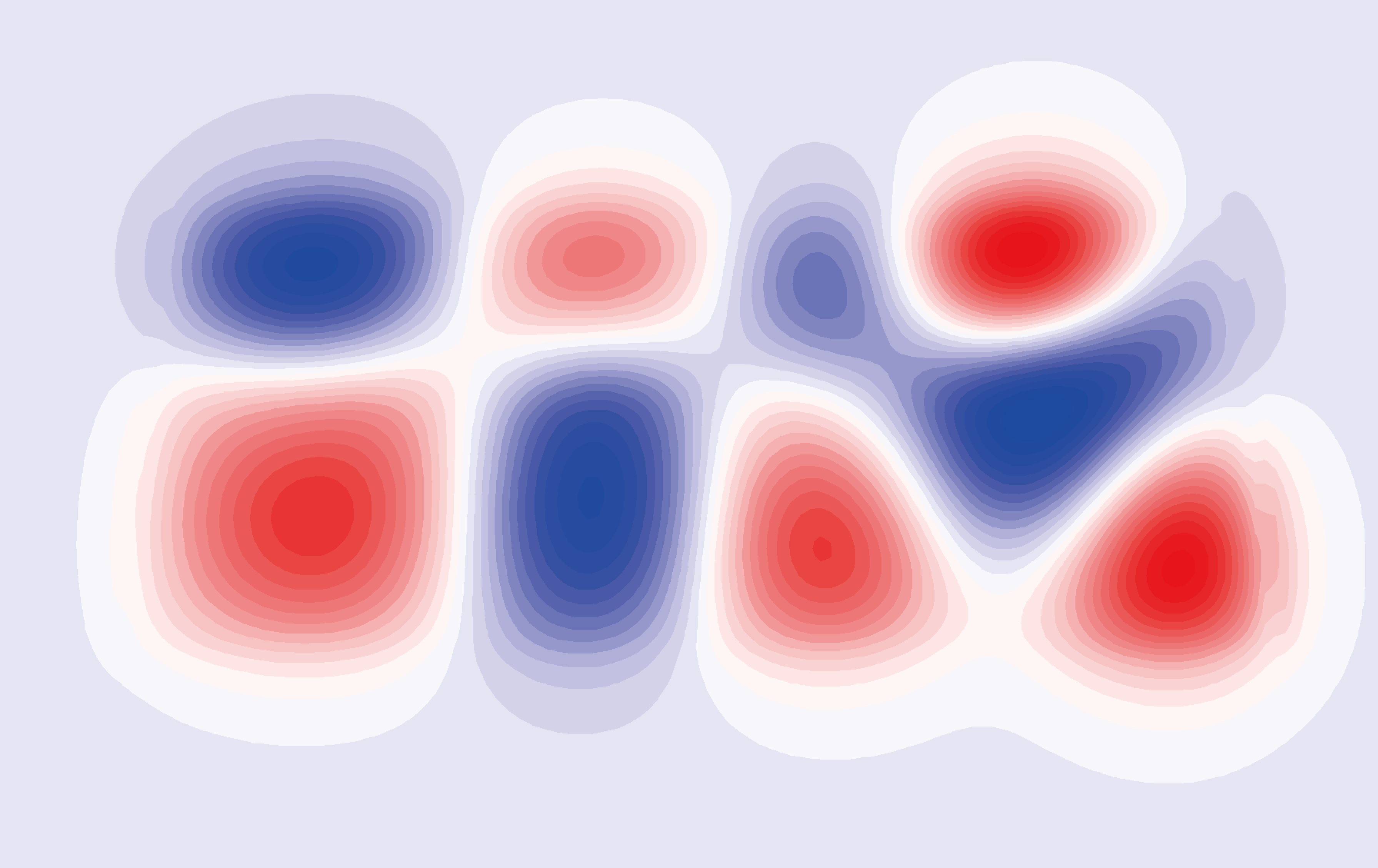} & \includegraphics[width=1.3cm]{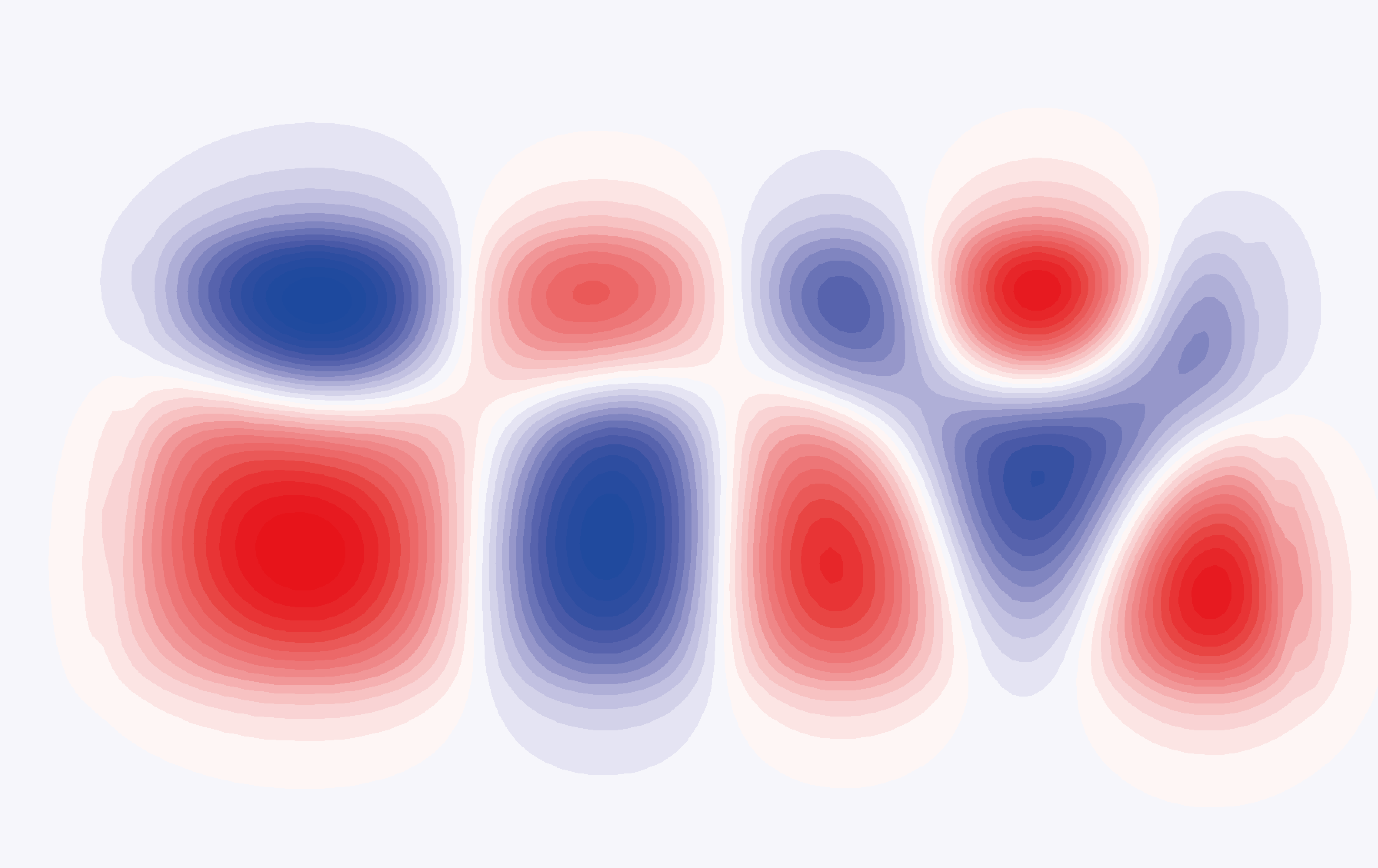}\\
Calculated Overlap Factor & 0.15 & 0.28 & 0.11 & 0.54 & 0.52 & 0.48 & 0.21 & 0.14\\
\hline
\end{tabular}
  \label{tab:modeprofile}
\end{table*}

Frequency comb generation has been the focus of many recent research efforts due to its viability in a myriad of applications such as atomic clocks \cite{Papp:14,Diddams:01}, frequency metrology \cite{Keilmann:04,Udem:02}, microwave generation \cite{Diddams:00,Yi:15}, molecular spectroscopy \cite{Haakestad:13,Cruz:15} and ultrafast optics \cite{Cundiff:03,Roslund:14}. Integrated photonic structures are widely used for comb generation because of their compact nature and their ability to enhance nonlinear effects using resonators with small mode volumes. Silicon nitride $\mathrm{(Si_3N_4)}$ and aluminum nitride $\mathrm{(AlN)}$ in particular have been popular choices of material platform due to their wide bandgap, low loss wave-guiding and attractive nonlinear optical properties \cite{Guo:16,Okawachi:11,Jung:13,Liu:14,Pfeiffer:17}. Owing to their high inherent cubic optical nonlinearity, broadband octave combs in these materials have been realized \cite{Li:17,liu:16}. Furthermore, stable soliton generation in $\mathrm{(Si_3N_4)}$ has been demonstrated \cite{Li:17,Brasch:16,Pfeiffer:17} in the recent years. However, to be utilized in many of the aforementioned applications, a frequency comb must have its carrier-envelope offset (CEO) frequency stabilized \cite{delhaye:08,Bartels:09}. One approach is to employ 2f-3f interferometry  \cite{Brasch:17,Wang:16} to implement self-referencing a microcomb of lesser than an octave span. In this scheme, highly efficient on-chip second-harmonic generation (SHG) and third-harmonic gneration (THG) are desirable. 

Among the many methods for THG \cite{Kajzar:85,Lippitz:05,New:67,Hong:13,Levy:11,Carmon:07}, doubly-resonant cavities are preferred because the nonlinear optical effects are resonance enhanced while significantly reducing the need for pump power. Despite efforts to fully integrate THG sources on-chip using various platforms such as photonic crystal cavities \cite{Corcoran:09}, germanium nano-disks \cite{Grinblat:16} and silicon nitride microring resonators \cite{Levy:11}, challenges remain to achieve a high THG efficiency. We summarize that current approaches are limited by the following factors: (i) inefficiencies in the TH light extraction process, because of the large wavelength difference between IR input light and TH light. (ii) the trade-off between phase-matching and mode overlap. Due to the large wavelength difference, the fundamental modes for IR light must be phase-matched to higher order modes of TH light which leads to a lower modal overlap factor.

In this paper, we propose a platform that co-integrates $\mathrm{AlN}$ and $\mathrm{Si_3N_4}$, introducing a degree of freedom that allows for the precise tuning of the phase-matching condition and mode overlap, and will therefore enable highly efficient THG on-chip. Additionally, $\mathrm{AlN}$ is unique in that it can be sputter deposited at low temperature (<$350\thinspace{}^{\circ}\mathrm{C}$) on a variety of substrate materials including $\mathrm{Si_{3}N_{4}}$ while maintaining low loss and excellent nonlinear optical properties \cite{Pernice:12, Xiong:12, Pernice2:12}, allowing for flexibility in fabrication. Guided by our theoretical analysis and optimization of the modal overlap between the TH modes and the fundamental mode in the composite $\mathrm{AlN/Si_{3}N_{4}}$ structure, we experimentally realized microring resonators with small mode volume ($20\thinspace\mu\mathrm{m}$ radius) and high quality ($Q$) factor. Along with an optimized visible light extraction waveguide, we achieved, to the best of our knowledge an unprecedented on-chip THG efficiency of $180\%\thinspace\mathrm{W^{-2}}$.

This composite $\mathrm{AlN/Si_3N_4}$ microring can also be optimized for Kerr comb generation and SHG by utilizing the $\chi^{(2)}$ nonlinearity of AlN, providing a unique photonic platform for potential stabilized soliton generation by simultaneous Kerr comb generation and on-chip 2f-3f inteferometry. 

\section{System and Design}
\label{sec:system}

In this work, we demonstrate a hybrid $\mathrm{AlN}$ on $\mathrm{Si_{3}N_{4}}$ microring resonator to realize high efficiency THG. A high $Q$-factor microring resonator was fabricated to ensure a large resonance enhancement of the fundamental pump light (telecom) and the third-harmonic (TH) light. The cross-sectional geometry of the composite $\mathrm{AlN/Si_3N_4}$ microring and two coupling waveguides (for pump and TH light) were designed such that phase matching and a large modal overlap between the pump and TH modes were achieved. 

In a uniform microring resonator, optical modes satisfying the cylindrical symmetry and the orbit angular momentum is a good quantum number. Thus, the THG process in a microring requires the conservation of momentum, i.e. $3m_{a}=m_{b}$ with subscript $a$ and $b$ denoting the fundamental and TH modes. With this condition satisfied, we can describe the THG process by the following Hamiltonian
\begin{equation}
\begin{split}
H=\omega_{a}a^{\dagger}a+\omega_{b}b^{\dagger}b+g[(a^{\dagger})^{3}b+a^{3}b^{\dagger}] \\
+\varepsilon_{\mathrm{p}}(ae^{i\omega_{\mathrm{p}}t}+a^{\dagger}e^{-i\omega_{\mathrm{p}}t}).
\end{split}
\label{eq:hamiltonian}
\end{equation}
where $a(b)$ corresponds to the bosonic operator for the fundamental(TH) mode in the microring. $\omega_{a}$ and $\omega_{b}$ are the respective angular frequencies of the two modes. The pump field and frequency are respectively given by $\varepsilon_{\mathrm{p}}=\sqrt{2k_{a,1}\mathrm{P}_{\mathrm{p}}/\hbar\omega_{\mathrm{p}}}$ and $\omega_{\mathrm{p}}$. 
\begin{figure}[t]
\centering
\includegraphics[width=\linewidth]{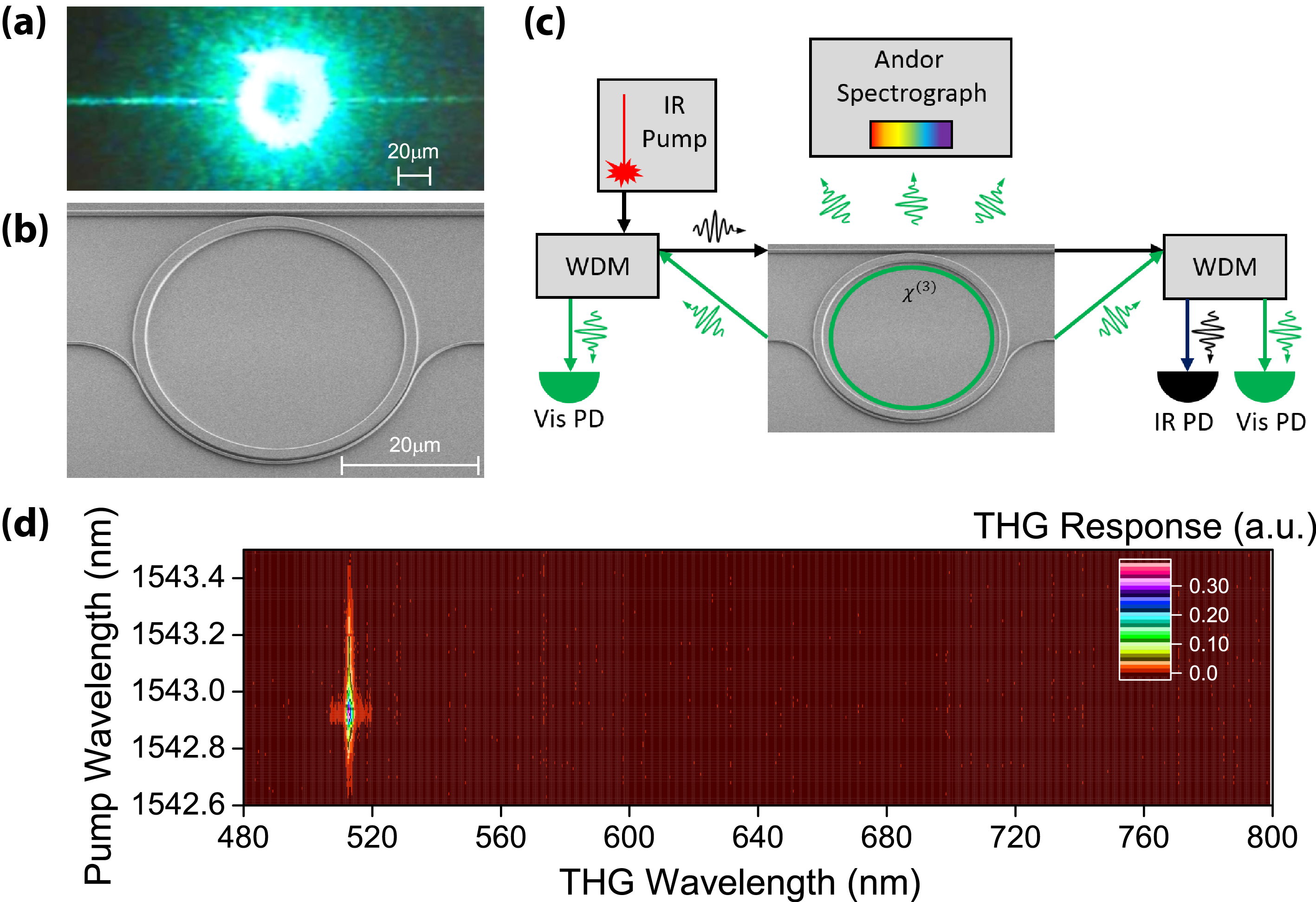}
\caption{\textbf{(a)} SEM image of a ring resonator device. \textbf{(b)} Top view of the TH light scattered from the ring. \textbf{(c)} Block diagram of the experimental setup including the Andor system situated on top of the ring resonator to collect scattered light. An IR continuous-wave pump source was amplified by an erbium-doped fiber amplifier (EDFA) and coupled into the ring resonator through a wavelength division multiplexer (WDM). Two visible photodetectors and one IR photodetector was used to monitor the TH output and transmission respectively. \textbf{(d)} Third harmonic generation spectrum with respect to input pump wavelength. Data was taken using the Andor Spectrograph 193i. At each pump wavelength input, TH light scattered from the ring was collected by the spectrograph with a slit opening of $50\thinspace\mu \mathrm{m}$ and an integration time of one second. The linewidth of the THG wavelength is due to the limit in resolution of the Andor Spectrograph.}
\label{fig:setup}
\end{figure}
The doubly resonant enhancement of the THG requires that the mode frequencies match each other as $3\omega_{a}\approx\omega_{b}$. Additionally, we also would like to optimize the nonlinear coupling strength
\begin{equation}
g\approx\frac{\zeta\sqrt{3}\hbar\omega_{a}^{2}}{\varepsilon_{0}2\pi R}\frac{\chi^{(3)}(r)}{\sqrt{\varepsilon_{a}^{3}\varepsilon_{b}}}\delta(m_{b}-3m_{a})
\label{eq:g}
\end{equation}
which is proportional to the modal overlap factor
\begin{equation}
\zeta=\frac{\int\int drdz[u_{a,z}^{*}(r,z)]^{3}u_{b,z}(r,z)}{\left[\int\int drdz|u_{a,z}(r,z)^{2}|\right]^{3/2}\left[\int\int drdz|u_{b,z}(r,z)^{2}|\right]^{1/2}}.
\label{eq:zeta}
\end{equation}
Here, $\chi^{(3)}(\textbf{r})$ denotes the third-order nonlinear susceptibility of the medium. $\varepsilon_{0}$ is the vacuum permittivity and $u_{a(b),z}(r,z)e^{im_{a(b)}\theta}$ is the electric field distribution of the optical modes within the microring. $\varepsilon_{a(b)}$ is the average relative permittivity of the microring for IR(visible) light. $R$ represents the average radius of the micro-ring. The nonlinear coupling strength $g$ is an important metric because the THG efficiency at critical coupling and zero detuning can be expressed as
\begin{equation}
\eta=g^{2}\frac{16Q_{a,0}^{3}Q_{b,0}}{\hbar^{2}\omega_{a}{}^{6}}.
\label{eq:efficiency}
\end{equation}
Here we see that the efficiency is proportional to $g^{2}$ and the quality factors $Q_{a,0}^{3}, Q_{b,0}$ of both the fundamental IR and the higher order TH mode.

From the analysis above, two conditions must be met. (1) The effective indices of the IR and visible modes must be the same in order to ensure momentum conservation, also referred to as the phase matching condition. (2) Since the efficiency scales with $\zeta^{2}$, a large field overlap can greatly enhance THG. In our design, the diameter of a typical ring resonator was 40$\,\mu \mathrm{m}$, with a fixed hybrid structure of $330\thinspace\mathrm{nm}$ sputtered polycrystalline AlN on $330\thinspace\mathrm{nm}$ of low pressure chemical vapor deposition (LPCVD) grown $\mathrm{Si_{3}N_{4}}$. The main degrees of freedom of this study used to tune into the phase matching condition as well as engineer the large field overlap were the hybrid layer design and ring width. A graph of the effective index of the high order TH modes (dotted lines) and the fundamental pump transverse electric (TE) mode (solid red line) with increasing ring width is plotted in Fig. \ref{fig:effindex}(a), which was calculated by commercial software FIMMWAVE. 

The process of optimizing THG of our devices started with systematically studying each phase-matched point by calculating the corresponding mode profile of the TH modes and the overlapping factor $\zeta$.  This was repeated for each intersecting point in the effective index plot between the fundamental IR TE mode and the high order visible TE modes. We found that the highest field overlap occurs when the fundamental IR TE mode is phase matched with the visible $\mathrm{TE_{02}}$ mode. Table \ref{tab:modeprofile} shows the simulated mode profiles for each phase matched width and its corresponding $\zeta$. Based on these theoretical results, we determined that the optimal width for efficient THG is at a ring width of $1.68\thinspace\mu \mathrm{m}$ and $1.86\thinspace\mu \mathrm{m}$. We experimentally observed optimal THG at a ring width of $1.85\thinspace\mu \mathrm{m}$.

Due to a large difference in wavelength the width of coupling waveguides differ for telecom light versus visible light. For a given waveguide geometry, the coupling strength of visible light from the ring to bus waveguide designed for IR light coupling is much weaker \cite{Guo2:16}, leading to an adversely affected overall efficiency. Therefore, we employed a wrap-around extraction waveguide for visible light to maximize our on-chip THG efficiency. The width of the extraction waveguide was tapered from 0.17 to 0.12$\thinspace\mu \mathrm{m}$ to provide phase matching for a wider range of TH light. The gap between the extraction waveguide and the microring was varied between $0.3-0.5\thinspace \mu \mathrm{m}$, with the highest extraction efficiency experimentally observed at $0.3\thinspace\mu\mathrm{m}$.

\section{Results}

An optical image of the hybrid $\mathrm{AlN/Si_{3}N_{4}}$ microring device during operation is shown in Fig.$\,$\ref{fig:setup}(a). At the cross section of the ring and waveguide, the height of both layers were fixed to be $0.33\thinspace\mu \mathrm{m}$. The $\mathrm{AlN/Si_{3}N_{4}}$ layers were grown on top of a $3.3\thinspace\mu \mathrm{m}$ thick silicon dioxide. The top point-contact bus waveguide and the bottom wrap-around extraction waveguide were used for the coupling of the input telecom light and the TH light respectively as displayed in Fig.$\,$\ref{fig:setup}(b-c). The design was realized by first using electron beam lithography to define the pattern of the devices. Then the $\mathrm{AlN}$ layer was dry etched using a $\mathrm{BCl_{3}/Cl_{2}/Ar}$ recipe, followed by the $\mathrm{Si_{3}N_{4}}$ layer etched under a $\mathrm{CHF_{3}/O_{2}}$ chemistry. Finally, a layer of silicon dioxide was deposited on top using plasma-enhanced chemical vapor deposition (PECVD) as a top cladding layer. A detailed diagram of the cross section is shown in Fig.$\,$\ref{fig:effindex}(b). 

The Q-factor of the device is crucial, as indicated by Eq.$\,$(\ref{eq:efficiency}), and was improved by minimizing both the scattering and absorption losses of the material. Under the $\mathrm{BCl_{3}/Cl_{2}/Ar}$ chemistry, AlN etches three times faster than $\mathrm{Si_{3}N_{4}}$. Therefore, a two step dry etching processes was developed with finely adjusted etch times to create the smoothest sidewall possible. Additionally, the devices were annealed at $\mathrm{950}\thinspace^{\mathrm{\circ}}\mathrm{C}$ for 1 hour, after which the intrinsic Q of the IR optical modes were improved by a factor of around two to an observed value of $Q_{a,0}\thickapprox6\times10^{5}$. 

For our experimental setup as shown in Fig.$\,$\ref{fig:setup}(c), at the input end is a tunable telecom laser source as well as a visible wavelength detector. Optionally, the tunable laser signal was amplified using an erbium-doped fiber amplifier. At the output end there are two detectors for telecom and visible light respectively. Additionally, wavelength division multiplexers were employed to seperate TH light from the input IR pump. Lastly, directly situated above the ring resonator device is the input to the Andor 193i Spectrograph and an iVac camera. The Andor spectrograph was calibrated to have a resolution of $\mathrm{0.15\thinspace nm}$.

Fig.$\,$\ref{fig:setup}(d) is a colormap indicating the relative intensity of the scattered TH light that was collected by the Andor spectrograph through a lens placed directly on top of the working device. A slit size of $50\thinspace\mu \mathrm{m}$ and an integrating time of one second was used. The relative intensity was plotted with respect to the pump wavelength and the output wavelength. We observe from the scattered light, an intense radiation at the TH wavelength of $\mathrm{514\thinspace nm}$, and no light at the second harmonic wavelength (which should be observed at around $771\thinspace\mathrm{nm}$. This verifies that the light is not generated by cascaded second-order nonlinear processes, where a sum-frequency between the pump and second harmonic light can give rise to emission at the TH wavelength.

\begin{figure}[ht]
\centering
\includegraphics[width=\linewidth]{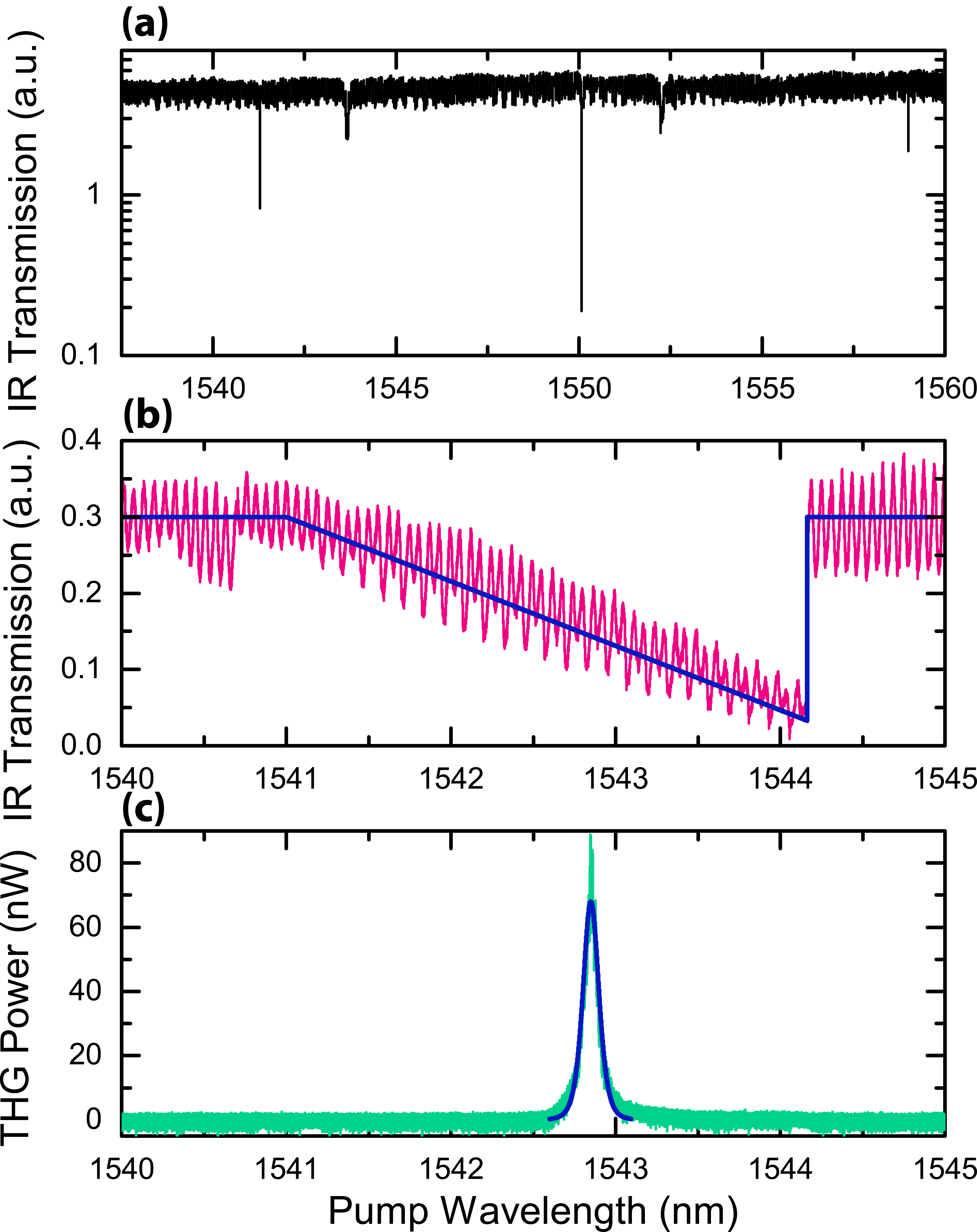}
\caption{\textbf{(a)} Transmission at low input pump power.  The resonance within the shaded region was used for generating TH light. \textbf{(b)} Transmission at high power of the IR pump resonance. Red curve and blue curve denote the experimentally observed and the simulated resonance respectively. The triangular shape resonance is a result of the thermal shifting of the $1541.2\thinspace\mathrm{nm}$ resonance at high temperatures. The large shift is due to high confinement of thermal energy within the waveguide. \textbf{(c)} The TH response with respect to the pump wavelength. The response (green) is modeled by a lorentzian shaped curve (blue).}
\label{fig:transmission}
\end{figure}
Our microring device was characterized by sweeping the pump laser across the resonance at input powers of $230\thinspace\mathrm{mW}$ and observing the third harmonic response as seen in Fig.$\,$\ref{fig:transmission}(a-c). Due to the thermal effect, at high pump powers the microring temperature increases when the laser wavelength approaches the resonance and more light is absorbed, which causes a shift of resonance wavelength \cite{Carmon:04}. Here we assumed that the shift is dominated by thermal effects, and can be modeled by the coupled differential equations describing the dynamics of temperature and cavity field. In the classical limitation where optical fluctuations are neglected such that $a=\alpha$ and at steady state, the coupled equations read
\begin{eqnarray}
\frac{d}{dt}\alpha=-i[(\omega_{a}-\omega_{\mathrm{p}})-\kappa_{a}-c\Delta T]\alpha-i\kappa_{a,1}\epsilon_{\mathrm{p}}=0,\\
\frac{d}{dt}\Delta T=K|\alpha|^{2}-\gamma_{th}\Delta T=0.
\end{eqnarray}
Here $\Delta T$ is the temperature difference between the microring cavity and the surrounding environment, $K$ is the heating coefficient describing the change in temperature, and $c$ describes the thermal nonlinear coefficient of mode $a$.

Due to the balancing between heating and thermal dissipation described by thermal relaxation time $\gamma_{th}^{-1}$, the temperature and cavity frequency thermal shift is dependent on the input pump intensity and detuning. Comparing Fig.$\,$\ref{fig:transmission}(a) and (b), one is the transmission curve of low input power whereas the other has $|\mathrm{P_p}|^{2}\approx230\thinspace\mathrm{mW}$ (an on-chip pump power of around $60\,\mathrm{mW}$), the thermal effect gives rise to a triangular shaped transmission curved with a sudden jump corresponding to the thermal bistability. Solving for the equations above, we arrive at a theoretical fitting that agrees well with the pump transmission in Fig.$\,$\ref{fig:transmission}(b). Furthermore, the change in temperature ($T$) induced by the pump could also shift the wavelength of other modes in the micoring, due to the thermal-optical change of refractive index $\Delta n \propto T$. As a result, the thermal frequency shift for different wavelengths is different ($\Delta \omega/\omega \approx -\Delta n$). So, when sweeping the pump light $\omega_{\mathrm{p}}$, the power absorbed by the microring and $\Delta T$ changes, as the frequency mismatching between pump and TH modes $3\omega_{\mathrm{p}}-\omega_a (T)$ approaches $0$ at a specific pump detuning. As shown in Figs.$\,$\ref{fig:transmission}(c), the maximum THG is achieved as the pump wavelength approaches the IR resonance, at a cavity absorbed power of approximately $30\,\mathrm{mW}$.  The intracavity photon then gives rise to a considerable thermal shift. From our experimental measurements, we fitted the resonance thermo-shift by $\lambda_{a}(T)=\lambda_{a0}+d_{a}T$, $\lambda_{b}(T)=\lambda_{b0}+d_{b}T$, with the coefficients as $d_{a}=0.016\pm0.001\,\mathrm{nm/K}$, $d_{b}=0.0046\pm0.0003\,\mathrm{nm/K}$.

\begin{figure}[ht]
\centering
\includegraphics[width=0.7\linewidth]{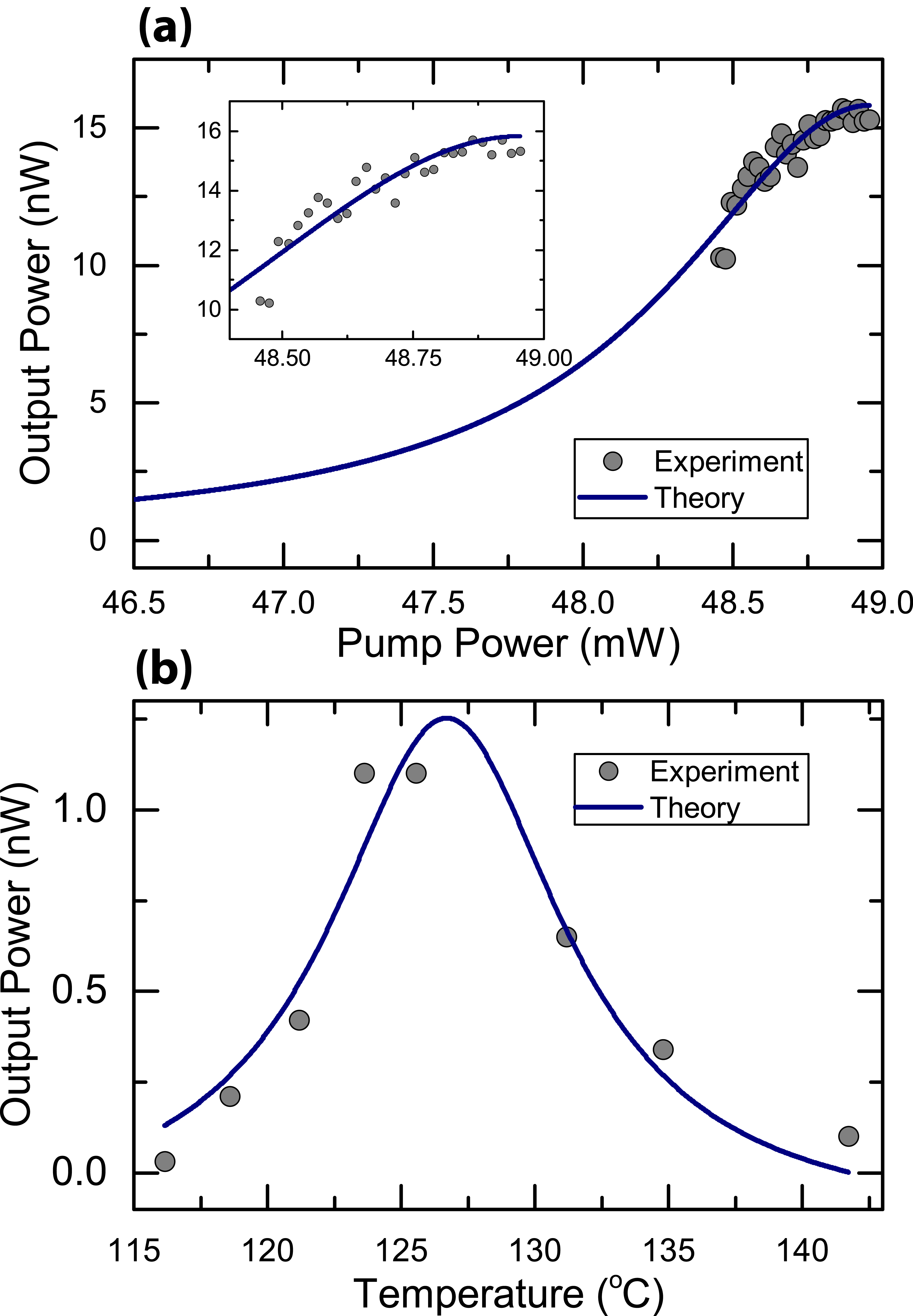}
\caption{\textbf{(a)} Study of the THG power dependence using a continuous-wave (CW) laser input. The horizontal axis shows the amount of pump power that is coupled into the resonance at the phase-matched wavelength. The solid curve indicates the model prediction of the output power at high input powers, where a difference in resonance shift speeds causes resonance wavelength mismatch and subsequently a deviation from a cubic power dependence.  \textbf{(b)} Temperature dependence of the THG. The temperature was tuned by coupling a (CW) input into a resonance away from the wavelength at which THG occurs. The absolute temperature was then calibrated for by measuring the thermal shift in the IR resonances. Simultaneously, a low-duty cycle modulated laser with a average power of $4\thinspace\mathrm{mW}$ was used for THG. This scheme was implemented to avoid the thermal shift at high CW input powers.  A lorentzian shaped temperature dependence curve with a full-width-at-half-maximum (FWHM) of $10.6^{\circ}\mathrm{C}$ is fitted.}
\label{fig:powerdepend}
\end{figure}

Including the thermal effects, the theoretical expression for on-chip THG power in waveguide is given by
\begin{equation}
\mathrm{P_{THG}}\approx\frac{2g^{2}\kappa_{b,1}\left(\frac{2\kappa_{a,1}}{\kappa_{a}^{2}}\right)^{3}\frac{\hbar\omega_{b}}{(\hbar\omega_{\mathrm{p}})^{3}}(\mathrm{P}_{\mathrm{p}})^{3}}{[\frac{2\pi c}{\lambda_{a}\lambda_{b}}((\lambda_{a0}-3\lambda_{b0})+(d_a-3d_b)T)]^{2}+\kappa_{b}^{2}}.
\label{eq:thgpower}
\end{equation}
From which we see that the power from THG has a cubic dependence on the pump power, and is also dependent on the wavelength mismatch. Hence the power-dependence curve is non-cubic, as shown in Fig.$\,$\ref{fig:powerdepend}(a). By optimizing the temperature, we observed a maximum THG output of $\mathrm{49\thinspace} \mu \mathrm{W}$ in waveguide at $30\mathrm{mW}$ absorbed power. This corresponds to an on-chip THG efficiency of $\eta=\mathrm{P_{THG}}/(\mathrm{P_{p}})^{3}=180\%\thinspace\mathrm{W^{-2}}$ and an absolute conversion efficiency of 0.16\%.

At $4\thinspace\mathrm{mW}$ average input power, we investigated the tunability of THG efficiency using temperature control as shown in Fig. \ref{fig:powerdepend}(b), where the maximum THG efficiency occurs at $\lambda_{a}-3\lambda_{b}=0$. The theoretical formula for the temperature dependence of the THG efficiency becomes
\begin{equation}
\eta =\frac{\mathrm{P_{THG}}}{\mathrm{P_{p}}^3} \approx \frac{\hbar\omega_{b}}{(\hbar\omega_{\mathrm{p}})^{3}} \frac{16g^{2}\kappa_{b,1}\kappa_{a,1}^3}{\kappa_{a}^6 [\frac{2\pi c}{\lambda_{a}\lambda_{b}}(d_a-3d_b)T]^{2}+\kappa_{b}^{2}]}.
\label{eq:thgpower2}
\end{equation}
It is clear from the data shown in Fig. \ref{fig:powerdepend}(b), that temperature dependence of the THG efficiency $\eta$ has a Lorentzian shape, and is well fitted by our theoretical expression, indicated by the solid line. The full-width at half maximum (FWHM) is determined by: $\Delta T=\frac{\kappa_{b}\lambda_{a}\lambda_{b}}{\pi c(d_a-3d_b)}=\frac{\lambda_{a}}{Q_{b}d_{\Delta}}$. Where $d_{\Delta} = (d_a-3d_b)$. From our experimental results, we observed $\Delta T=10.6\,\mathrm{^{\circ}C}$. Therefore, according to the previously measured value of $\lambda_{a}$ and $d_{\Delta}$, we estimate $Q_{b}\approx6.4\times10^{4}$. 
%
\section{Conclusion}

In conclusion, THG was achieved on a composite $\mathrm{AlN} / \mathrm{Si_{3}N_{4}}$ integrated platform. The structure of the ring was optimized after systematically studying the field overlap of the fundamental input mode and the higher order TH modes. From our experimental results, we report a calibrated on-chip efficiency of $\eta=\mathrm{P_{THG}}/(\mathrm{P_{p}})^{3}=180\%\thinspace\mathrm{W^{-2}}$. To the best of our knowledge, this is the highest reported THG efficiency achieved on an integrated platform. Additionally, a power dependent and temperature dependent measurement was used to confirm our theoretical understanding of the THG. We note that future improvements can be made to our system to achieve higher efficiency THG, including further optimization of the IR and visible quality factor, the thickness of the bilayer as well as the gap between the microring and the visible coupling waveguide.

\section*{Funding Information}
Defense Advanced Research Projects Agency (DARPA); David and Lucile Packard Foundation.

\section*{Acknowledgments}

Facilities used for device fabrication were supported by Yale SEAS cleanroom and Yale Institute for Nanoscience and Quantum Engineering. The authors thank Michael Power and Dr. Michael Rooks for assistance in device fabrication.

\bibliography{eff_thg}

\ifthenelse{\boolean{shortarticle}}{%
\clearpage
\bibliographyfullrefs{thgbib}
}{}

\end{document}